# Fully Coupled Electromagnetic-Thermal-Mechanical Comparative Simulation of Direct vs. Hybrid Microwave Sintering of $3Y\text{-}ZrO_2$

Charles Manière[a], Tony Zahrah[b], Eugene A. Olevsky[a, c]*

(a) Powder Technology Laboratory, San Diego State University, San Diego, USA
(b) Matsys Inc., Sterling, USA
(c) NanoEngineering, University of California, San Diego, La Jolla, USA

**Keywords**
Microwave Sintering, Finite Element Method, Densification, Direct Heating, Hybrid Heating, Zirconia

## Abstract

Direct and hybrid microwave sintering of $3Y\text{-}ZrO_2$ are comparatively studied at frequency of 2.45 GHz. Using the continuum theory of sintering, a fully coupled electromagnetic-thermal-mechanical (EMTM) finite element simulation is carried out to predict powder samples deformation during their microwave processing. Direct and hybrid heating configurations are computationally tested using advanced heat transfer simulation tools including the surface to surface thermal radiation boundary conditions and a numeric proportional–integral–derivative (PID) regulation. The developed modeling framework shows a good agreement of the calculation results with the known experimental data on the microwave sintering of $3Y\text{-}ZrO_2$ in terms of the densification kinetics. It is shown that the direct heating configuration renders highly hot spot effects resulting in non-homogenous densification causing processed specimen's final shape distortions. Compared to the direct heating, the hybrid heating configuration provides a reduction of the thermal inhomogeneity along with a densification homogenization. As a result of the hybrid heating, the total densification of the specimen is attained without specimen distortions. It is also shown that the reduction of the sample size has a stabilization effect on the temperature and relative density spatial distributions.

---

* Corresponding author: EO (ACerS Fellow): Powder Technology Laboratory, San Diego State University, 5500 Campanile Drive, San Diego, CA 92182-1323,
Ph.: (619)-594-6329; Fax: (619)-594-3599, *E-mail address*: eolevsky@mail.sdsu.edu

## I. Introduction

The microwave sintering process showed appealing results for the densification[1-8], synthesis[9], assembling[10], annealing[11] of various materials and for the achievement of enhanced materials properties[12-14]. A wide range of ceramic[15-17], metallic[18-23], polymeric[24-25], composite[26-27] material systems have been successfully fabricated by this technology. Compared to the conventional sintering (utilizing a convective indirect heating[28-30] pattern), a volume and direct microwave heating[1,3] of the sample is applicable with potentially high heating rates[31] of about 100K/min. This processing specifics results in energy savings and in the reduction of the grain growth allowing to produce materials with dense and fine microstructures[32-36] possessing high mechanical properties. Because of these benefits, the microwave sintering process encounters a growing number of applications in both research and industrial fields.

Many authors reported an acceleration of the sintering kinetics in microwave compared to conventionally sintered samples[37-41] for the same thermal cycle. The origin of this so-called "non-thermal effect" or "microwave effect" is still debated but a possible explanation can be the action of ponderomotive forces at the pore/grain boundary junctions[42-46] and/or diffusion mechanism enhancements induced by the microwave alternative fields[47-48]. In addition to the earlier mentioned benefits, this phenomenon also contributes to the reduction of the sintering time and promotes fine microstructures.

Despite all its benefits, microwave sintering remains a rather difficult process to control. The temperature and displacement measurements are not easy to perform because thermocouples or pushrods inserted in the furnace working space may interact with microwaves[49-50]. As a rule, non-contact measurements using pyrometer and/or camera are employed in microwave sintering with the calibration of the measurement devices as described by Croquesel et al.[37]. Another difficulty of this technology is the inherent heating instabilities[51-54]. The wave propagation in the microwave cavity intrinsically generates areas of high and low electric and magnetic fields intensity. Therefore, depending of the sample location, the heating can be very different and then non-reproducible[55]. In a previous work[56], we showed that the temperature-dependent evolution of the dielectric properties of ceramic materials can promote the formation of highly hot spots in the center of the sintered samples, thereby causing problems of inhomogeneous densification. This hot spot phenomenon appears often under conditions of the direct microwave heating. To stimulate the temperature homogenization and to assist the processing of low dielectric loss materials, a hybrid heating using susceptors can be used. A susceptor is a material highly sensible to the microwaves. Silicon carbide is often used for susceptor components since this material exhibits high values of dielectric loss at low temperatures[57-60]. Numerous researches[21,61-64] reported a more stable heating using susceptors in hybrid or indirect heating configurations. In order to aid the adjustments of the tooling to a specific microwave sintering application, a respective simulation tool is needed.

The finite element (FE) simulation is a powerful tool for the prediction of:

- the areas of high electric/magnetic fields intensity in the microwave cavity;
- the thermal gradients generation;
- the sample densification gradient and deformation for simple or complex shapes;
- the process regulation parameters;
- the effect of susceptors on the hybrid heating of the sample.

The FE simulation allows optimization studies for the heating/densification stabilization. Microwave sintering simulations encompass three main physics highly coupled to each other: electromagnetism, heat transfer and sintering. In the past, researchers studied the electromagnetic-thermal (EMT) aspects of the microwave sintering process using the finite element[61-62,64,65-67], or finite-difference time-domain[68-69] (FDTD) methods. In 2010, Bouvard et al.[64] included the sintering aspect of the process through an empirical densification equation and coupled the densification to the other physics assuming a fixed specimen's geometry during the sintering cycle. Birnboim et al.[70] and Riedel et al.[71] used similar approaches to conduct coupled simulations of microwave sintering. More recently, Abedinzadeh et al.[72] modeled a pressure assisted microwave cavity. They studied first the cold compaction by the Gurson–Tvergaard–Needleman model[73-76] and then the microwave heating was modeled using COMSOL Multiphysics™[77], and finally they investigated the hot densification based on the previously obtained cold pressing and heating modeling data.

Overall, fully or partially coupled EMTM models exist in the literature, but they are rare and sparse. In most cases, these studies employ simplified model conditions like idealized sample geometry; material parameters, which are not always temperature/porosity dependent; and idealized conduction and radiation heat exchange conditions.

In the present study, a fully coupled EMTM simulation of pressureless microwave sintering is investigated using COMSOL Multiphysics™ software. The purpose of this modeling approach is to include the coupled effects of the electromagnetic fields, heating, densification and complex deformation of a powder sample in the comparative analysis of the direct and hybrid heating taking into account the influence of the sample dimensions. In this study, the susceptor technique is explored as this approach is simple and highly adaptable to a wide range of microwave furnace. Below, a $TE_{102}$ resonant rectangular waveguide cavity is investigated at the excitation frequency of 2.45GHz. In the developed model the dielectric, thermal and mechanical parameters are coupled to each other and evolve with the temperature and porosity. The surface to surface thermal radiation between all the parts and PID regulation are also taken into account.

## II. Theory and calculations

II.1. EMTM model description

II.1.1 Governing equations

The microwave propagation in the cavity can be modeled by the following combination of the Maxwell's equations:[77] .

$$\nabla \times (\mu_r^{-1} \nabla \times \boldsymbol{E_r}) = k_0^2 \left(\varepsilon_r - \frac{j\sigma}{\omega\, \varepsilon_0}\right) \boldsymbol{E_r} \tag{1}$$

with $\mu_r$ being the complex relative permeability, $\varepsilon_r$ the complex relative permittivity, $\sigma$ the electric conductivity, $k_0$ the vacuum wave number, $\omega$ the angular frequency, $\varepsilon_0$ the vacuum permittivity, $j$ the complex number, and $\boldsymbol{E_r}$ is defined by the harmonic electric field expression $\boldsymbol{E} = \boldsymbol{E_r} exp(j\omega t)$.

The heat transfer part of the model is described by the heat and electromagnetic loss equations:

$$\rho C_p \frac{\partial T}{\partial t} + \nabla.(-\kappa \nabla T) = Q_e \tag{2}$$

$$Q_e = \frac{1}{2}(\varepsilon_0 \varepsilon_r'' \boldsymbol{E}^2 + \mu_0 \mu_r'' \boldsymbol{H}^2) \tag{3}$$

where $\rho$ is the density, $\kappa$ is the thermal conductivity, $C_p$ is the specific heat, $\mu_0$ the vacuum permeability, $\boldsymbol{H}$ is the magnetic field intensity, $\varepsilon_r''$ and $\mu_r''$ the permittivity and permeability imaginary parts, respectively.

The heat losses depend on the electric and magnetic fields intensity and on the values of the permeability and permittivity imaginary parts which evolve with the temperature and with the relative density of the sample and/or of the tooling.

The densification part of the EMTM model is based on the continuum theory of sintering[78]. The rheological basis of this model has been initially developed by Skorokhod[79]. In its present form the continuum theory of sintering[78] is able to predict sintering utilizing both pressure[80-82] or pressureless[83-84] methods; and it is based on the consideration of the viscoplastic (nonlinear-viscous) porous material behavior.

In accord with the continuum theory of sintering[78], the stress tensor in a porous material is defined by:

$$\underline{\sigma} = \frac{\sigma_{eq}}{\dot{\varepsilon}_{eq}}\left(\varphi\underline{\dot{\varepsilon}} + \left(\psi - \frac{1}{3}\varphi\right) tr\left(\underline{\dot{\varepsilon}}\right)\mathbb{i}\right) + Pl\mathbb{i} \quad (4)$$

where $\sigma_{eq}$ and $\dot{\varepsilon}_{eq}$ are the equivalent stress and strain rate, $\underline{\dot{\varepsilon}}$ is the strain rate tensor, $\mathbb{i}$ is the identity tensor, $Pl$ is the sintering stress, $\varphi$ and $\psi$ are the shear and bulk moduli defined as functions of the porosity $\theta$:

$$\varphi = (1-\theta)^2 \quad (5)$$

$$\psi = \frac{2}{3}\frac{(1-\theta)^3}{\theta} \quad (6)$$

The sintering stress expression depends on the average particle radius $r_0$, on the porosity, and on the surface energy $\alpha$:

$$Pl = \frac{3\alpha}{r_0}(1-\theta)^2 \quad (7)$$

The equivalent strain rate is defined as:

$$\dot{\varepsilon}_{eq} = \frac{1}{\sqrt{1-\theta}}\sqrt{\varphi\dot{\gamma}^2 + \psi\dot{e}^2} \quad (8)$$

with the strain rate tensor invariants:

$$\dot{e} = \dot{\varepsilon}_x + \dot{\varepsilon}_y + \dot{\varepsilon}_z \quad (9)$$

$$\dot{\gamma} = \sqrt{2\left(\dot{\varepsilon}_{xy}^2 + \dot{\varepsilon}_{xz}^2 + \dot{\varepsilon}_{yz}^2\right) + \frac{2}{3}\left(\dot{\varepsilon}_x^2 + \dot{\varepsilon}_y^2 + \dot{\varepsilon}_z^2\right) - \frac{2}{3}\left(\dot{\varepsilon}_x\dot{\varepsilon}_y + \dot{\varepsilon}_x\dot{\varepsilon}_z + \dot{\varepsilon}_y\dot{\varepsilon}_z\right)} \quad (10)$$

The equivalent stress and strain rate are related by a power law creep equation:

$$\dot{\varepsilon}_{eq} = A(T)\sigma_{eq}^n \quad (11)$$

where *A* is a parameter defining the decrease of the material resistance to deformation with the temperature. The stress exponent *n* defines the nonlinearity of the material behavior. Usually the free sintering experiments show linear viscous behavior (n=1), with the exception of the nano-powders that possess high values of sintering stress and nonlinear[85] behavior (n>1). For the linear viscous material behavior, *A(T)* represents the reciprocal of the material viscosity $\frac{1}{2\eta(T)}$.

### II.1.2 Boundary conditions

The microwave propagation in the cavity implies two main boundary conditions: the "port" and the totally reflective wall conditions. The port condition describes an inlet of the microwave power in a predefined propagation mode. The aim of the metallic wall of the cavity is to reflect and contain the wave inside the cavity. Commonly they are modeled by totally reflective wall conditions or directly by assuming a metal material of the wall.

The main thermal boundary conditions describe thermal radiation. For the external surface radiating towards the surface under ambient temperature, the "surface to ambient" conditions can be used:

$$\varphi_{rsa} = \sigma_s \epsilon ({T_e}^4 - {T_a}^4) \qquad (12)$$

with $\varphi_{rsa}$ being the radiative heat flux, $\epsilon$ the emissivity, $\sigma_s$ the Stefan–Boltzmann's constant, $T_e$ and $T_a$ the emission and ambient temperature, respectively.

For the surfaces that radiate towards each other, the total outgoing radiative heat flux $J$ (called radiosity) is defined as the sum of the wall thermal radiation $\epsilon e_b(T)$ and the reflected part (*refl*) of the outside incoming irradiation G (see fig. 1).

$$J = refl + \epsilon e_b(T) = (1 - \epsilon)G + \epsilon (Nr)^2 \sigma_s T^4 \qquad (13)$$

with *Nr* being the refractive index. The net inward heat flux expression $\varphi_{rss}$ is then:

$$\varphi_{rss} = \epsilon (G - e_b(T)) \qquad (14)$$

The "surface to surface radiation" boundary condition considers each points radiating in every direction.

The boundary conditions of the mechanical part of the problem (the sintering) are prescribing point displacements (u, v, w) on the sample bottom face.

## II.2. Simulation configurations and material properties

In this work three direct and hybrid heating configurations are explored. A rectangular waveguide cavity is investigated in $TE_{102}$ mode (fig. 2a). The tooling includes a high temperature insulating box[86] (maximum operating temperature 2073 K) made of alumina 80%-silica 20% and a silicon carbide susceptor ring for the hybrid heating (see fig. 2b). The sample is a cylindrical green compact of 3Y- $ZrO_2$ with an initial relative density of 55%. The first configuration is the direct heating of a 12 mm diameter sample, the second configuration is similar but uses a smaller (7 mm diameter) sample and the last is a hybrid heating configuration with the 12 mm diameter sample (with SiC susceptor). For each samples the height is the same of the diameter. Contrarily to flat samples, these high height samples geometry have enough volume space to allow the creation of high hot spot and allow to see the distortion they generate. The aim of these three configurations is to study the impact of the microwave heating on the densification, temperature gradients, scale factor for the direct heating configuration and the potential stabilization of the hybrid heating configuration.

The material properties for 3Y-$ZrO_2$ are represented in Table.1. These properties depend on both the temperature (T) and relative density (D). The sintering process conditions based on the experimental data of Wroe and Rowley[41] (see fig. 3) render a good agreement of the experimental and the developed model data in terms of the densification kinetics. The alumina-silicate and silicon carbide temperature dependent properties are reported in Table. 2. The temperature cycle is similar to the reference cycle[41] (10 K/min from 300 to 1673 K and 20 min of dwell at 1673 K) and is imposed at the point at the center of the upper sample face. A numeric PID regulation is employed to utilize this thermal cycle regulating the input power $P_{in}$:

$$P_{in} = K_p e(t) + K_I \int_0^t e(t) d\tau + K_D \frac{de(t)}{dt} \quad (15)$$

where $e(t)$ is the temperature error (Tcycle - Tregulation), $K_p$, $K_I$, $K_D$, are the proportional, integral, and derivative coefficients, respectively.

For the direct heating, we determined the PID coefficients: $K_p = 20$, $K_I = 4$, $K_D = 1$. For the hybrid heating the behavior of the heating changes and the following PID coefficients are adapted: $K_p = 10$, $K_I = 0.05$, $K_D = 150$. Because the hybrid heating configuration provides a long thermal response between the imposed electromagnetic power input and the sample heating, the derivative coefficient is high ($K_D = 150$) which slows down the regulation and helps avoiding the PID regulation runaway.

## III. Modeling Results

III.1. Electromagnetic wave propagation in the cavity.

Depending on the waveguide shape and dimensions, different propagation modes can be applied[90-92]. The dimensions of the considered rectangular waveguide (fig. 2a) have been calculated using Eq. (16) to obtain a $TE_{102}$ resonant mode[93] for the excitation frequency of 2.45 GHz:

$$\omega_{nmp} = c\sqrt{\left(\frac{n\pi}{a}\right)^2 + \left(\frac{m\pi}{b}\right)^2 + \left(\frac{p\pi}{l}\right)^2} \qquad (16)$$

where *a*, *b* are the rectangular cross-section dimensions, *l* is the waveguide length, *c* is the light velocity, n, m, p are the mode coefficients (1; 0; 2 for $TE_{102}$), and $\omega_{nmp}$ is the resonant frequency. Typically, the $TE_{102}$ mode in empty cavity possesses two principal maximum areas of the electric field as shown in fig.4a. In order to allow the maximum heating efficiency, the 3Y-$ZrO_2$ sample is placed in one of these areas[37]. Because the real part of 3Y-$ZrO_2$ permittivity is equal roughly to 10, the electric field is higher outside the sample in the direct (figs. 4b,c) and hybrid (fig. 4d) configurations. The disturbances of the electric field caused by the dielectric properties of 3Y-$ZrO_2$ generate a displacement of the maximum electric field point location in the cavity. The sample location is then adjusted to be in the highest electric field area among the different configurations. The hybrid heating configuration (fig. 4d) renders the maximum field disturbance due to the presence of both the sample and the susceptor. Because the sample is a good dissipative material only at high temperatures, the input power in high at low temperatures and about 2000W and fall to about 200W at high temperatures.

III.2. Heating and thermal exchanges

The temperature profiles of the tooling and sample at end of the heating cycle are reported in fig. 5 for the three configurations. The black lines represent the initial geometry. For these configurations, the insulation box succeeds to decrease the temperature of the insulated bottom face being in contact with the other support tools (Tmax = 800K). The sample lateral surface radiation is responsible for the insulation box inner lateral faces' temperature rise (up to 1400 K for the large dimension sample (fig. 5a) and 900 K for the small sample dimension (fig 5b)). The first direct heating case shown in fig. 5a points out to the presence of the two main sample/insulation heat fluxes: one is the thermal conduction at the sample bottom face of about 1E5 $W/m^2$ and the other is the surface to surface radiation at the other sample faces of about 2.6E5 $W/m^2$. The radiative flux is higher because the sample radiates heat to a wide colder insulation box surface. On the contrary, the conduction flux is limited by the insulation box low density and its high low thermal conductivity. The surface integral of the incoming/emitted radiative flux and the surface integral of the total net inward radiative flux and conductive flux on the sample bottom face are reported in fig.6 for the three configurations.

As a first step, the comparison of the incoming/emitted flux on the radiative surfaces shows that the incoming radiation flux from the heated insulating box is not negligible for the direct heating configurations (figs. 6a and 6b) and represents roughly 30% of the incoming flux. The total net inward radiative flux on the sample surfaces for the direct configurations is therefore negative (heat loss) and about 150 W for the larger sample and 75 W for the smaller sample. For the hybrid configuration, the SiC susceptor produces high amount of heat that radiates the sample (fig. 6c). For the hybrid configuration, the incoming radiation flux is of the same order of magnitude as the one of the sample producing this radiation flux. Consequently, the emitted and incoming radiation compensate each other and the total surface inward radiation flux is low and is about 30 W. In our case the hybrid heating configuration acts as a “thermal insulation” of the sample surfaces.

On the other hand, the surface integral of the conductive heat flux at the sample/insulation box interface is very low with the value of about 15 W for the direct configurations (figs. 6a and 6b) and 5 W for the hybrid configuration (fig. 6c). As discussed earlier, the main reason of these low values is the very low thermal conductivity of the insulating box, but the main explanation of the difference between the total inward surface radiation values (conduction of 15W vs. radiation of 150W) is the difference in the interface area, which is 5 time lower than the total surface area for thermal radiation. Consequently, the thermal insulation condition for this contact interface with the support tool is a good approximation, similarly to the previously obtained results[56].

III.3. Densification/heating comparison

The relative density and temperature fields for the 12 mm diameter sample direct configuration are reported in fig. 7. The temperature and relative density kinetic curves at the sample upper surface and center indicate the formation of a hot spot at the center of the sample with maximum temperature differences ($\Delta T_{surf\text{-}center}$) of about 500 K while the surface maximum temperature ($\Delta Tmax\text{-}_{surf}$) is between 20 and 100 K. This $\Delta T_{surf\text{-}center}$ difference represents two regimes: at the first stage $\Delta T_{surf\text{-}center}$ increases exponentially, at the second stage, when the densification occurs, the temperature difference $\Delta T_{surf\text{-}center}$ stops increasing and maintains the value of about 500 K up till the end of the cycle. As it was shown in a previous work[56], the hot spot formation is due to the $\varepsilon_r''$ increase with the temperature and the reduction of the size of the sample in turn reduces the hot spot phenomenon. Additionally, because $\varepsilon_r$ increases with the temperature and relative density, the penetration depth $d_p$ given by eq. (17) evolves from meters to millimeters for high temperatures and relative density (see fig. 8).

$$d_p = \frac{c}{2\pi f\sqrt{2\varepsilon_r'\left(\sqrt{1+\left(\frac{\varepsilon_r''}{\varepsilon_r'}\right)^2}-1\right)}} \quad (17)$$

where $f$ is the frequency, $c$ is the light speed and $\varepsilon_r'$ is the real part of the relative permittivity.

Then at roughly 1300K the penetration depth is of the same order of magnitude as the sample dimensions. Above this temperature, the electromagnetic heating starts to become more superficial and the thermal gradient stops increasing.

Moreover, considering the presence of the hot spot and taking into account that the penetration depth value is around 5 mm when the densification happens (see figs. 7 and 8), one can see that the area of electromagnetic heat loss follows the densification front from the center to the edge of the sample. Afterwards, when densification happens, the temperature gradients are stabilized.

Concerning the sample shape evolution (fig. 7), the hot spot generates a non-cylindrical shape due to a heterogeneous densification field. The densification starts from the center and propagates to the sample's edge. The sample first assumes a highly non-cylindrical shape and goes back to the cylindrical configuration at the end of the sintering. The final shape (fig. 7) is close to a cylinder except the corners remain not fully densified due to lower temperatures at these locations.

The 7 mm sample direct heating experiment (fig. 9) shows similar results except the temperature difference $\Delta T_{surf\text{-}center}$ is reduced to 300 K. Even despite this value is still generally high, the value of $\Delta T_{surf\text{-}center}$ is now sufficiently low to allow a complete densification of all the sample areas. This reduction of the temperature gradient can be explained by the size of the sample being of the same order of magnitude as the size of the hot spot area in the previous configuration (fig. 7). Therefore, the smaller the sample is, the better the temperature and densification homogeneity is.

For the large sample configuration, a technical solution enabling the reduction of the thermal gradients is needed. The hybrid heating configuration is studied for this purpose.

As shown in fig. 6, the hybrid configuration balances the sample thermal radiation by an incoming radiation of the same order of magnitude. It is claimed in the literature that the hot spot formation at the center of the sample can be balanced by the susceptor heating[1-3, 62, 64-65]. Indeed, our present simulation (fig. 10) shows a substantial degree of the homogenization of the temperatures in the radial direction (of about 100 K.) In the vertical direction, the temperature difference is reduced to about 200 K, but the radiation at the upper face, which is less subjected to the susceptor radiative compensation, is less intensive. Nevertheless, the total densification of the larger sample becomes possible with rendering a cylindrical final shape.

## IV. Conclusions

A fully coupled electromagnetic-thermal-mechanical simulation of the microwave sintering of $3Y\text{-}ZrO_2$ has been carried out. The developed modeling framework shows a good agreement of the calculation results with the known experimental data on microwave sintering of $3Y\text{-}ZrO_2$ in terms of the densification kinetics. Direct and hybrid microwave heating configurations have been tested for different sample dimensions. The direct heating configuration shows the formation of a hot spot at the center of the sample. The hot spot phenomenon increases drastically in the beginning of the process and then stabilizes when the densification occurs resulting at the end of the heating cycle in a temperature difference of 500 K across the sample’s volume. The microwave field penetration and then the heating become more superficial when the densification happens; this fact explains the temperature gradients’ stabilization. The sample shape is highly deformed during the densification due to the hot spot formation but tends to go back to the cylindrical shape at the end of the densification. Decrease of the sample size appears to reduce the thermal gradients and, at the same time, it resolves the problem of densification heterogeneities. This result shows that the low dimension samples' sintering is more stable. The hybrid heating configuration succeeds in reducing the temperature gradients in the sample allowing a more uniform overall densification of the large sample. Concerning the heat exchange aspects, the direct configuration surprisingly shows that the far-range radiation from the insulating tool is non-negligible in the surface-to-surface radiation heat exchanges. The use of SiC susceptors allows to balance the sample incoming amount of heat by offsetting it through thermal radiation.

## Acknowledgements

The authors gratefully acknowledge the support from the Office of Naval Research, Contract # N00014-14-C-0233, and Dr. William Mullins, Program Officer.

**Figure captions**

Fig. 1: Scheme of the surface to surface radiation exchanges at one surface point.

Fig. 2: a) Dimensions of the $TE_{102}$ rectangular waveguide cavity with the internal tooling and sample, b) Cutting slice of direct/hybrid tooling configurations (direct heating is the same configuration without the susceptor).

Fig. 3: Densification model used based on Wroe and Rowley [41], experiment/model comparison.

Fig. 4: Electric (multislice), magnetic (red arrows) cavity fields for: a) empty cavity, b) direct heating, c) direct heating small sample, d) hybrid sample heating configuration.

Fig. 5: Temperature field for the sample and tooling at the end of the heating cycle for: a) direct heating, b) direct heating small sample, c) hybrid sample heating configuration (the black lines represent the initial geometry).

Fig. 6: Surface integral of the incoming radiation (tooling→sample), produced thermal radiation (sample→tooling) and net inward surface thermal flux for: a) direct heating, b) direct heating small sample, c) hybrid sample heating configuration.

Fig. 7: Relative density (RD) and temperature fields for the direct heating configuration.

Fig. 8: Penetration depth temperature and relative density dependence; the red line indicates the sample dimensions (12 mm).

Fig. 9: Relative density (RD) and temperature fields for the direct small sample heating configuration.

Fig. 10: Relative density (RD) and temperature fields for the hybrid heating configuration.

**Table captions**

Table 1: Electromagnetic-thermal-mechanical properties of 3Y-ZrO2 (T is temperature in K and D is the relative density).

Table 2: Electromagnetic-thermal properties of the alumina-silica insulating box and silicon carbide susceptor (T is temperature in K).

Table 1: Electromagnetic-thermal-mechanical properties of 3Y-$ZrO_2$ (T is temperature in K and D is the relative density).

| | Temperature range (K) | Expression |
|---|---|---|
| $Cp$ [87] ($J.kg^{-1}.K^{-1}$) | 273-1473 | (43+2.35 T-0.34E-3 $T^2$+4.25E-6 $T^3$-2.09E-9 $T^4$+4.06E-13 $T^5$) × (1-1.5 × (1-D)) |
| | 1473-2200 | 638 × (1-1.5 × (1-D)) |
| $\kappa$ [87] ($W.m^{-1}.K^{-1}$) | 273-2200 | (1.96-2.32E-4 T+6.33E-7 $T^2$-1.91E-10 $T^3$) × (1-1.5 × (1-D)) |
| $\rho$ [87] ($kg.m^{-3}$) | 273-2200 | (6132 -9.23E-2 T-7.26E-5 $T^2$+4.58E-8 $T^3$-1.31E-11 $T^4$) × D |
| Emissivity ϵ | 273-2200 | 0.7 [88] |
| $\varepsilon_r'$ [56] | 273-2200 | -5.38-4.34E-3 T+2.22E1 D+1.37E-2 T D |
| $\varepsilon_r''$ [56] | 273-673 | 1.48E-1-5.76E-4 T-4.55E-01 D+1.77E-03 T D |
| | 673-873 | 3.82-6.03E-3 T-1.172E1 D+1.85E-2 T D |
| | 873-1073 | 1.56E1-1.95E-2 T-4.74E1 D+5.94E-2 T D |
| | 1073-2200 | 3.25E1-3.86E-2 T-7.64E1 D+8.46E-2 T D+3.82E-6 $T^2$+1.07 $D^2$ |
| $A = \frac{1}{2\eta}$ ($s^{-1}.Pa^{-1}$) | 273-2200 | (0.21/T) exp(-200000/(RT))<br>Identity from data ref [41] |

Table 2: Electromagnetic-thermal properties of the alumina-silica insulating box and silicon carbide susceptor (T is temperature in K).

| Material | | Temperature range (K) | Expression |
|---|---|---|---|
| $Al_2O_3$-$SiO_2$ | *Cp* [86] (J .kg$^{-1}$.K$^{-1}$) | 273-2200 | 1047 |
| | $\kappa$ [86] (W.m$^{-1}$.K$^{-1}$) | 273-2200 | 6.15E-2+1.74E-4 T |
| | $\rho$ [86] (kg .m$^{-3}$) | 273-2200 | 510 |
| SiC | *Cp* [87] (J .kg$^{-1}$.K$^{-1}$) | 273-673 | -8.35+3.08T-0.00293 $T^2$+1.0268E-6 $T^3$ |
| | | 673-1573 | 772+0.431 T-2.10E-5 $T^2$ |
| | | 1573-2200 | 1400 |
| | $\kappa$ [87] (W.m$^{-1}$.K$^{-1}$) | 273-2200 | 192-0.326 T+2.74E-4 $T^2$-7.71E-8 $T^3$ |
| | $\rho$ [87] (kg .m$^{-3}$) | 273-2200 | 2977+0.0510 T-2.29E-4 $T^2$+2.98E-7 $T^3$-1.92E-10 $T^4$+4.77E-14 $T^5$ |
| $Al_2O_3$-$SiO_2$ | Emissivity $\epsilon$ [89] | 273-2200 | 0.83 |
| SiC | | | 0.9 |
| SiC | $\varepsilon_r'$ [57] | 273-2200 | 1.88E-06 $T^2$-1.67E-03 T+6.4 |
| | $\varepsilon_r'$ [57] | 273-2200 | 2.36E-12 $T^4$-7.15E-09 $T^3$+7.72E-06 $T^2$-3.43E-03 T+9.92E-01 |
| $Al_2O_3$-$SiO_2$ | $\varepsilon_r$ | 273-2200 | 1 microwave transparent [61] |

F1

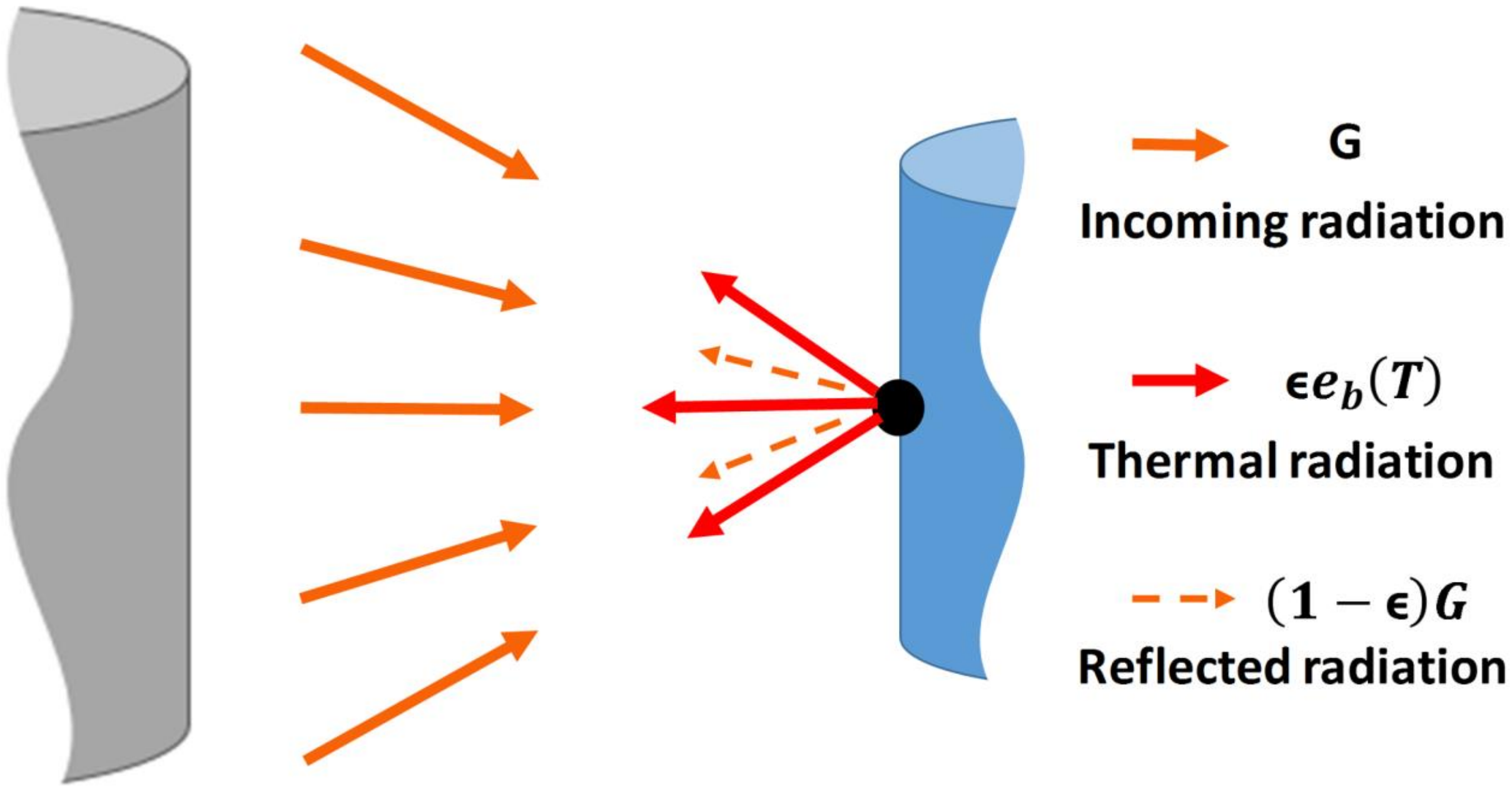

F2

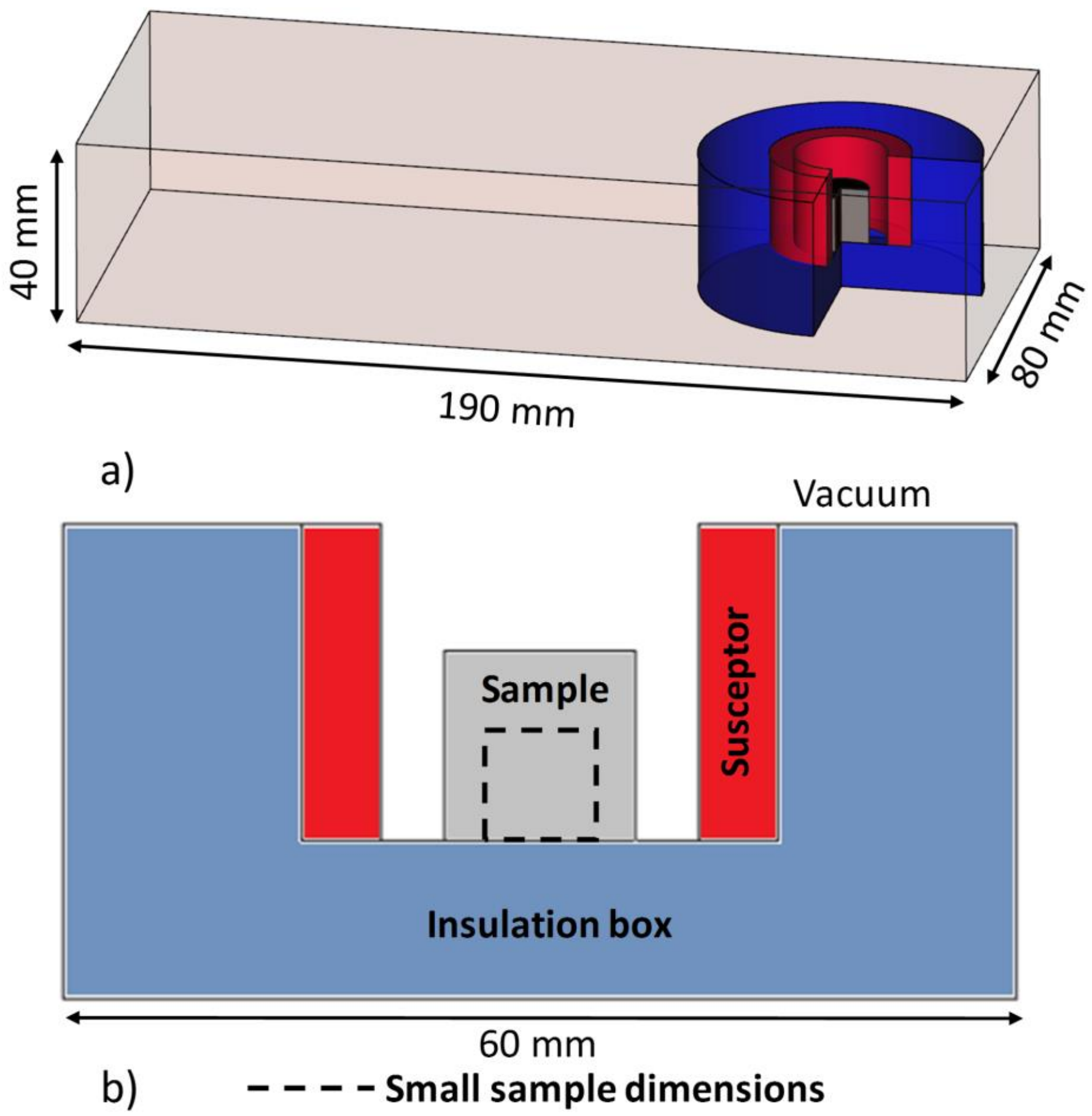

F3

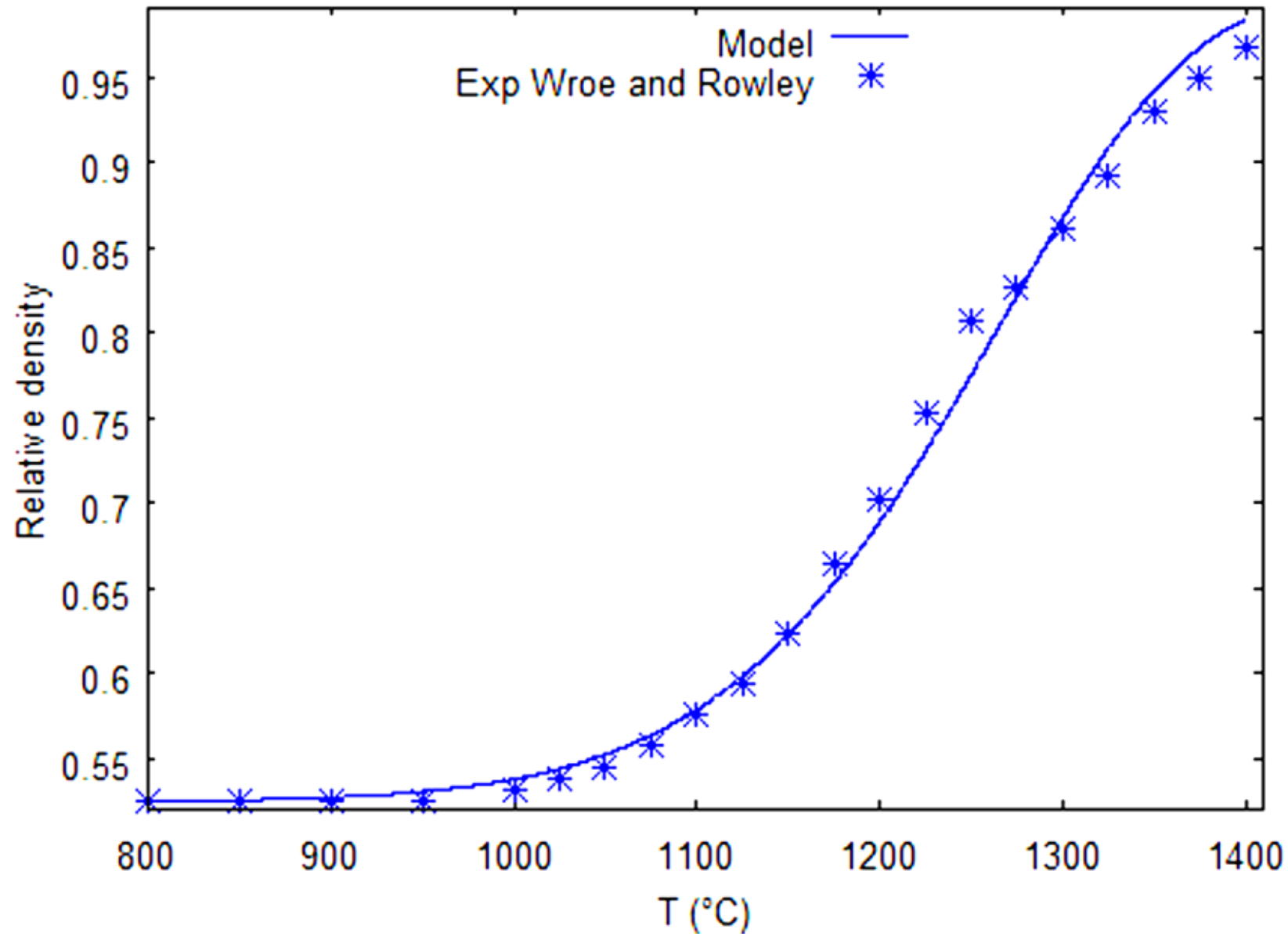

F4

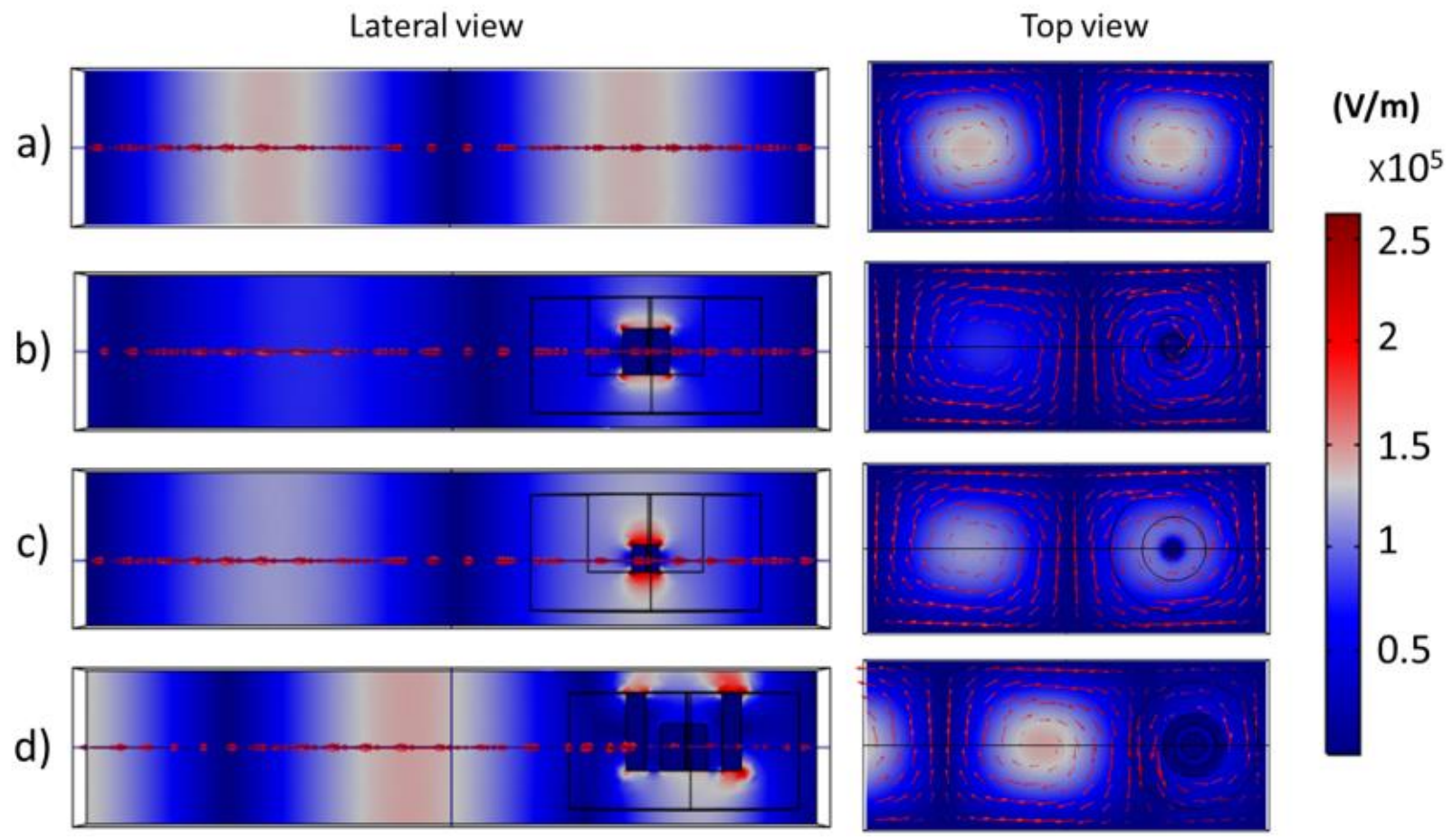

F5

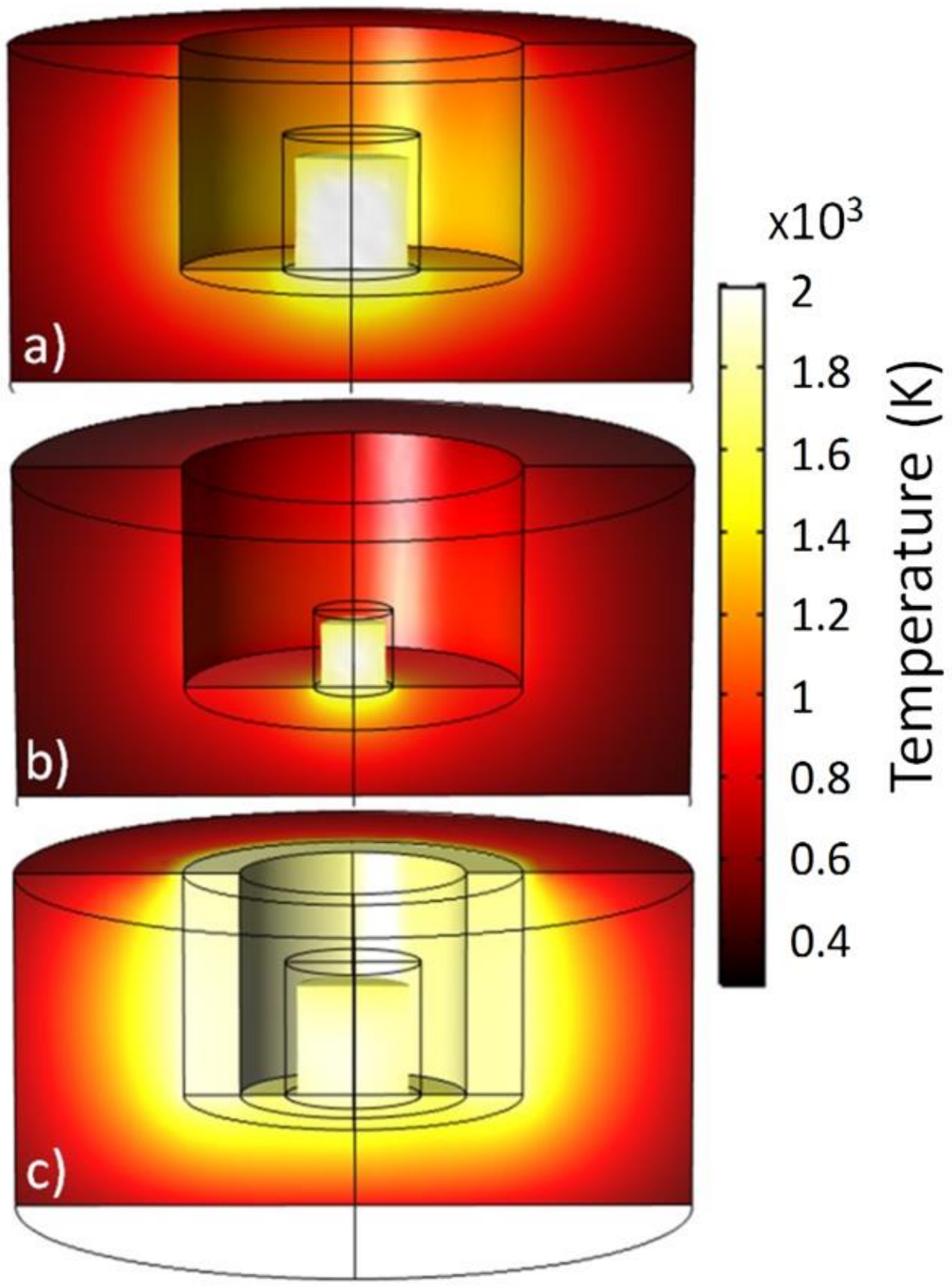

F6

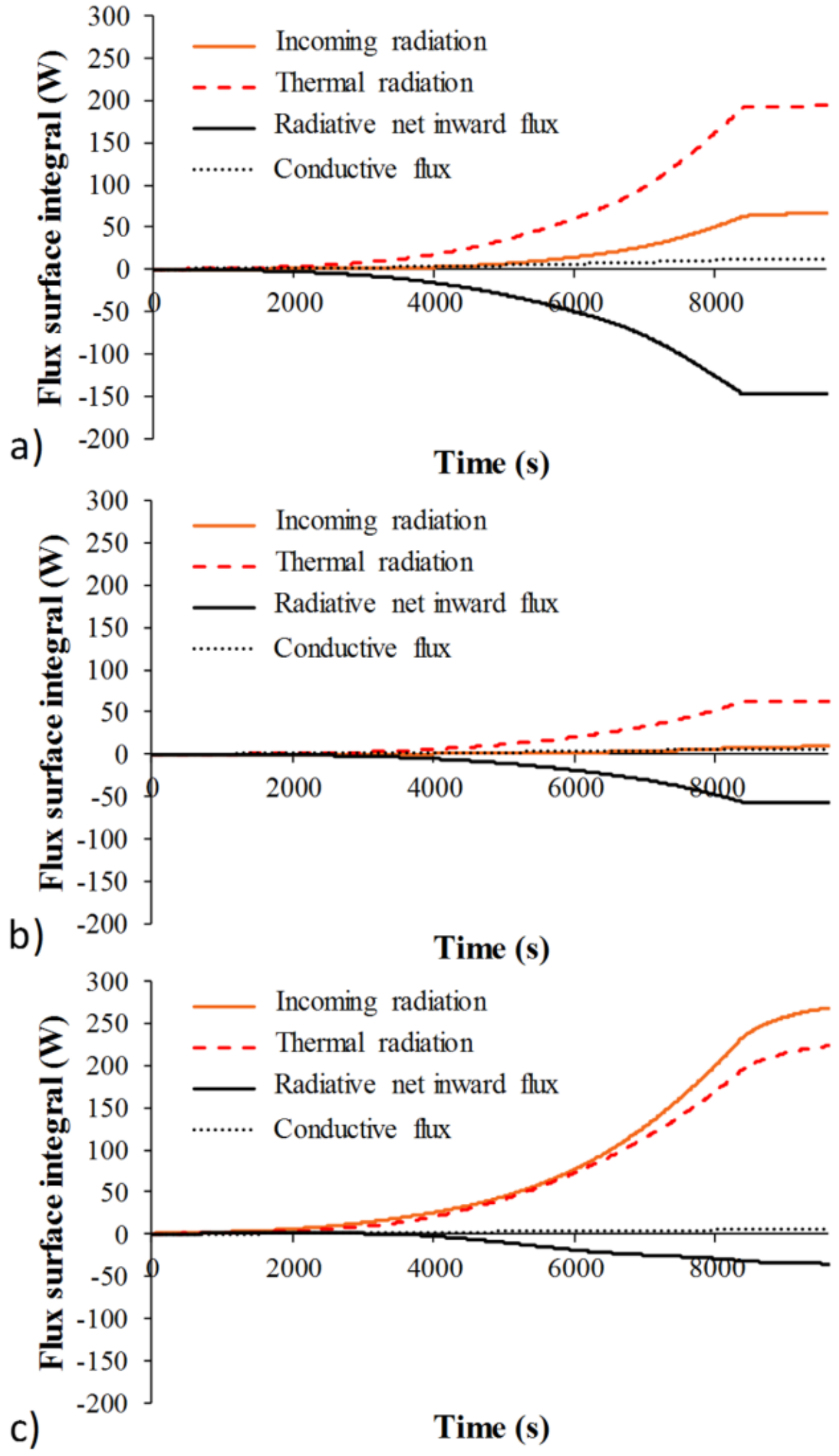

a)
Flux surface integral (W)
Incoming radiation
Thermal radiation
Radiative net inward flux
Conductive flux
Time (s)
b)
Flux surface integral (W)
Incoming radiation
Thermal radiation
Radiative net inward flux
Conductive flux
Time (s)
c)
Flux surface integral (W)
Incoming radiation
Thermal radiation
Radiative net inward flux
Conductive flux
Time (s)

F7

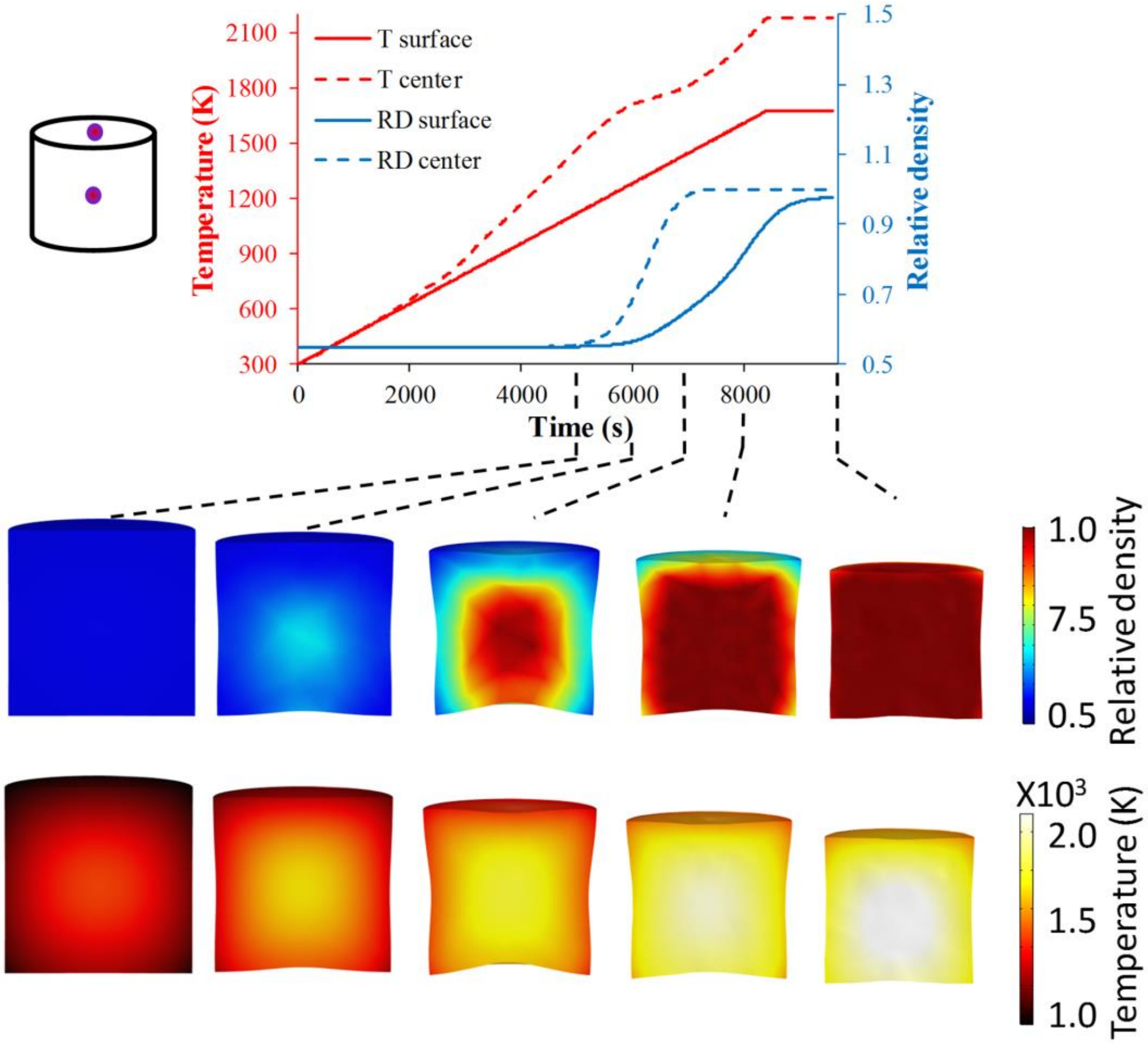

F8

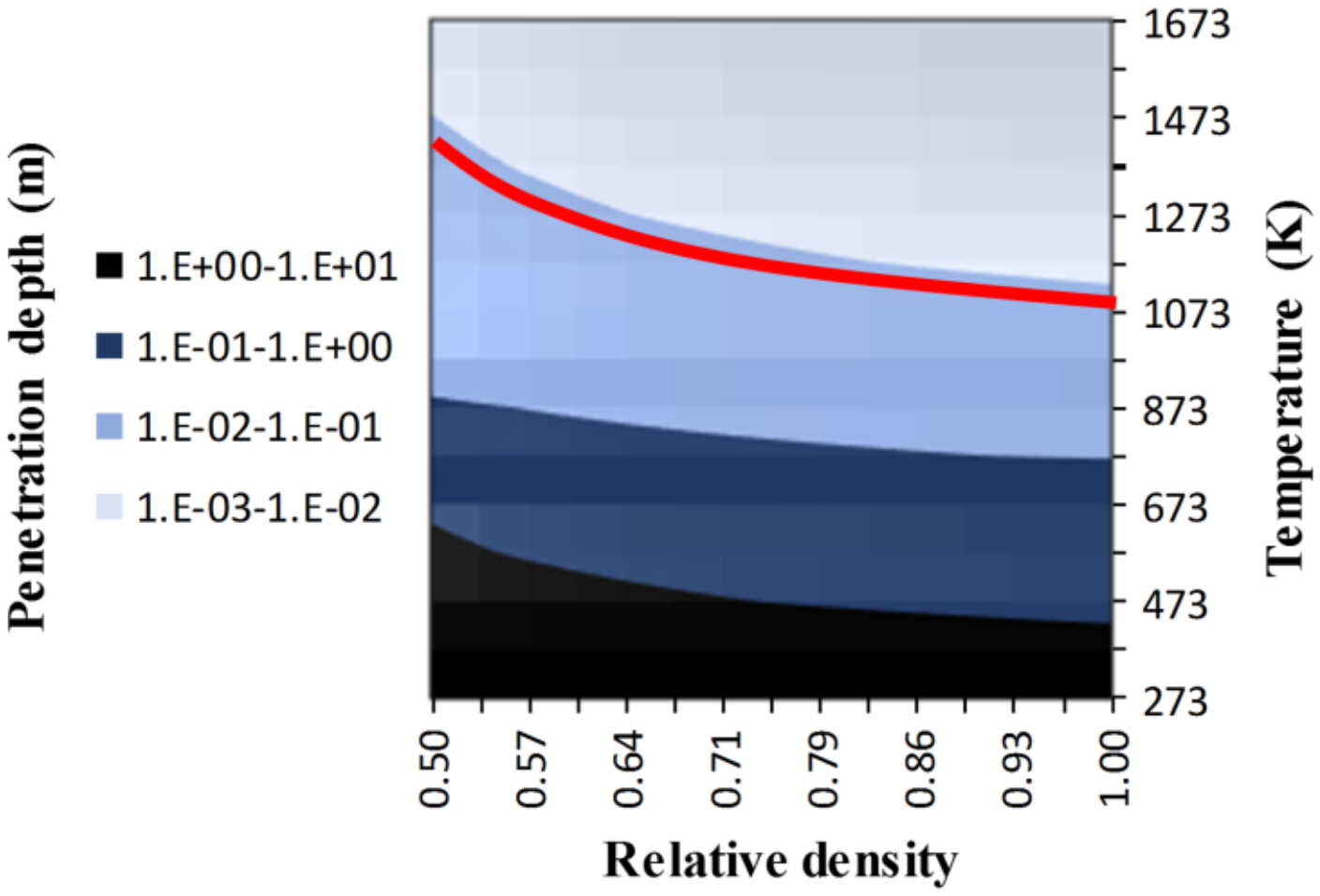

F9

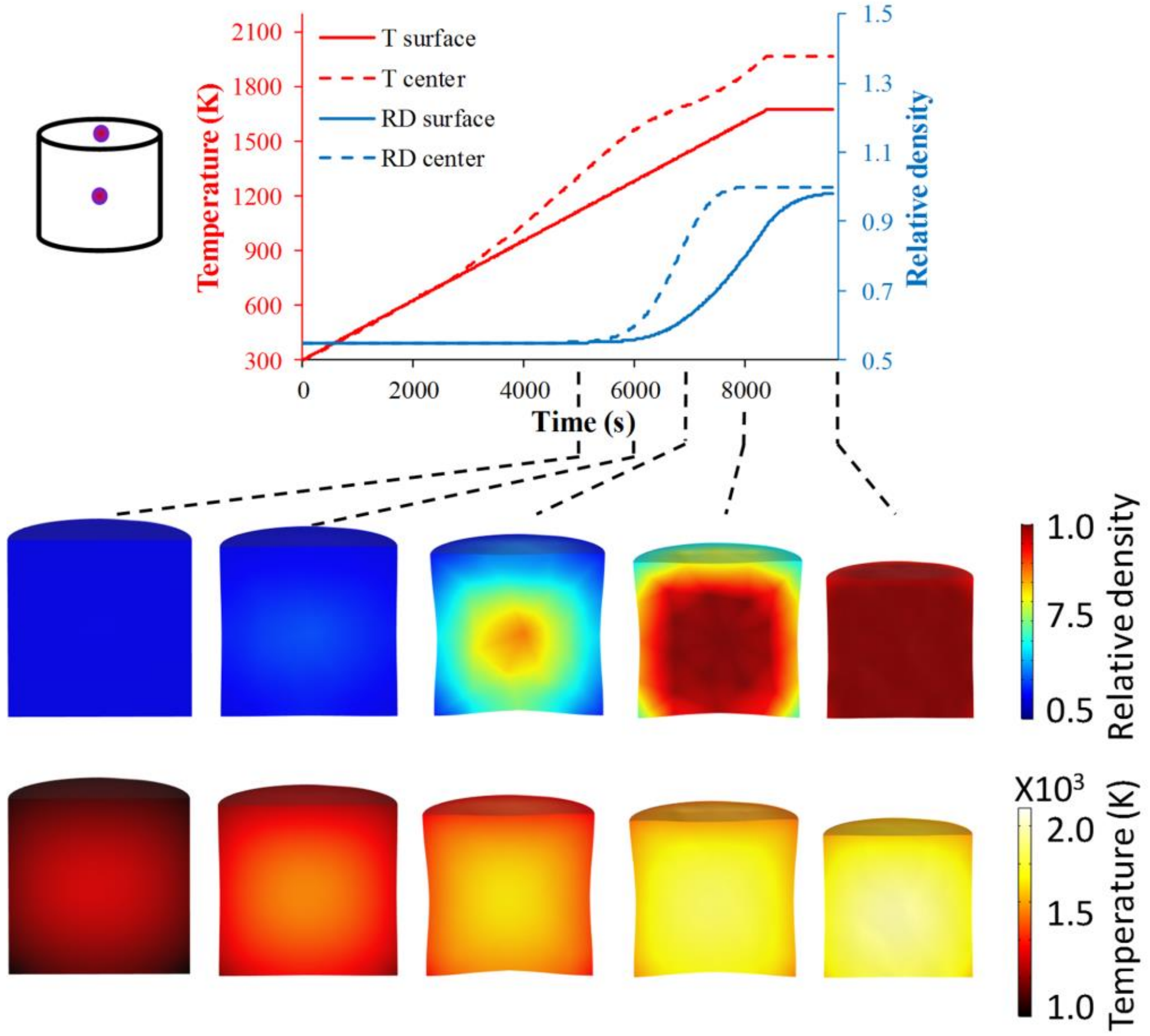

F10

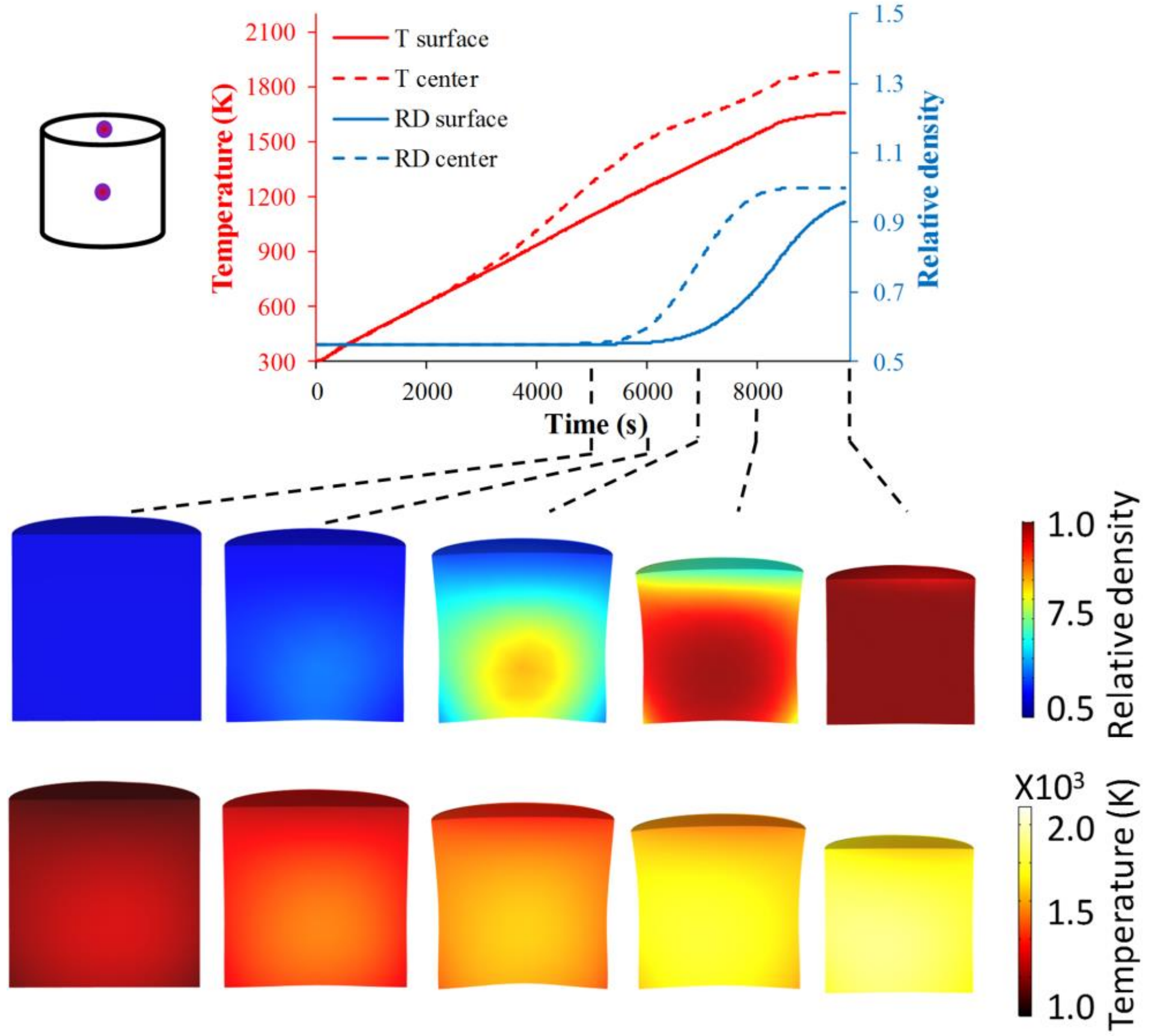